\documentclass[ALICE,manyauthors]{cernphprep}\newcommand{\publisher}{cern}
\usepackage{mathtools}
\usepackage{hyperref}

\newcommand{\pt}{\ensuremath{p_{\rm t}}}
\newcommand{\ptmin}{\ensuremath{p_{\rm t}^{\rm min}}}
\newcommand{\NA}{\ensuremath{N_{\rm A}}}
\newcommand{\kt}{\ensuremath{k_{\rm t}}}

\newcommand{\pp}{pp}
\newcommand{\PbPb}{Pb--Pb}
\renewcommand{\AA}{A--A}

\newcommand{\CKBNOTE}[1]{{\bf NOTE:  #1}}  
\renewcommand{\CKBNOTE}[1]{}  

\usepackage{graphics}
\usepackage{ifthen}
\usepackage{graphicx}
\usepackage{color}

\usepackage{amssymb}
\usepackage{multicol}
\usepackage{array}
\usepackage{caption}

\usepackage{subfigure} 

\usepackage[displaymath,modulo]{lineno}
\usepackage[english]{babel}



\ifthenelse{\equal{\publisher}{jhep}}{
  \title{Measurement of event background fluctuations for charged particle jet
  reconstruction in \PbPb\ collisions at $\sqrt{s_{\rm NN}} = 2.76$\,TeV
}

\author{The ALICE Collaboration}
  \collaborationImg{\includegraphics{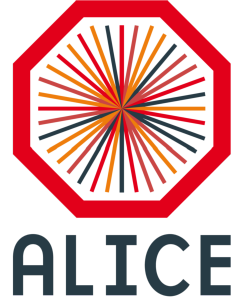}}
  \abstract{The effect of event background fluctuations on charged particle jet reconstruction in \PbPb\ collisions at $\sqrt{s_{_{\rm NN}}} = 2.76$\,TeV has been measured with the ALICE experiment. The main sources of non-statistical fluctuations are characterized based purely on experimental data with an unbiased method, as well as by using single high $\pt$ particles and simulated jets embedded into real \PbPb\ events and reconstructed with the anti-\kt\ jet finder. The influence of a low transverse momentum cut-off on particles used in the jet reconstruction is quantified by varying the minimum track $\pt$ between 0.15\,GeV/$c$ and 2\,GeV/$c$. For embedded jets reconstructed from charged particles with $\pt > 0.15$\,GeV/$c$, the
uncertainty in the reconstructed jet transverse momentum due to the heavy-ion background is measured to be 11.3\,GeV/$c$ (standard deviation) for the 10\% most central \PbPb\ collisions, slightly larger than the value of 11.0\,GeV/$c$ measured using the unbiased method. For a higher particle transverse momentum threshold of 2\,GeV/$c$, which will generate a stronger bias towards hard fragmentation in the jet finding process, the standard deviation of the fluctuations in the reconstructed jet transverse momentum is reduced to 4.8-5.0\,GeV/$c$ for the 10\% most central events. 
A non-Gaussian tail of the momentum uncertainty is observed and its impact on the reconstructed jet spectrum is evaluated for varying particle momentum thresholds, by folding the measured fluctuations with steeply falling spectra.

}
  \keywords{heavy-ion collisions, quark-Gluon plasma, hard scattering, underlying event fluctuations, jet reconstruction}
}

\begin{document}
\maketitle 
\ifthenelse{\equal{\publisher}{jhep}}{

}

\ifthenelse{\equal{\publisher}{cern}}{

    \PHnumber{2012-002}                 
    \PHdate{08 January 2012}              

  \begin{titlepage} 
    
    \title{Measurement of event background fluctuations for charged particle jet
  reconstruction in \PbPb\ collisions at $\sqrt{s_{\rm NN}} = 2.76$\,TeV
}


~

~

~

\ShortTitle{Background Fluctuations for Jet Reconstruction in \PbPb\ collisions}
\Collaboration{ALICE Collaboration \thanks{See Appendix~\ref{app:collab} for the list of collaboration 
                      members}}
\ShortAuthor{ALICE Collaboration}      



\begin{abstract}
The effect of event background fluctuations on charged particle jet reconstruction in \PbPb\ collisions at $\sqrt{s_{_{\rm NN}}} = 2.76$\,TeV has been measured with the ALICE experiment. The main sources of non-statistical fluctuations are characterized based purely on experimental data with an unbiased method, as well as by using single high $\pt$ particles and simulated jets embedded into real \PbPb\ events and reconstructed with the anti-\kt\ jet finder. The influence of a low transverse momentum cut-off on particles used in the jet reconstruction is quantified by varying the minimum track $\pt$ between 0.15\,GeV/$c$ and 2\,GeV/$c$. For embedded jets reconstructed from charged particles with $\pt > 0.15$\,GeV/$c$, the
uncertainty in the reconstructed jet transverse momentum due to the heavy-ion background is measured to be 11.3\,GeV/$c$ (standard deviation) for the 10\% most central \PbPb\ collisions, slightly larger than the value of 11.0\,GeV/$c$ measured using the unbiased method. For a higher particle transverse momentum threshold of 2\,GeV/$c$, which will generate a stronger bias towards hard fragmentation in the jet finding process, the standard deviation of the fluctuations in the reconstructed jet transverse momentum is reduced to 4.8-5.0\,GeV/$c$ for the 10\% most central events. 
A non-Gaussian tail of the momentum uncertainty is observed and its impact on the reconstructed jet spectrum is evaluated for varying particle momentum thresholds, by folding the measured fluctuations with steeply falling spectra.

\end{abstract}
\end{titlepage}
}

\ifthenelse{\equal{\publisher}{elsevier}}{

\title{Measurement of event background fluctuations for charged particle jet
  reconstruction in \PbPb\ collisions at $\sqrt{s_{\rm NN}} = 2.76$\,TeV
}

\begin{frontmatter}
\author[a1]{B.~Abelev et al.}\address[a1]{}
\begin{abstract}
The effect of event background fluctuations on charged particle jet reconstruction in \PbPb\ collisions at $\sqrt{s_{_{\rm NN}}} = 2.76$\,TeV has been measured with the ALICE experiment. The main sources of non-statistical fluctuations are characterized based purely on experimental data with an unbiased method, as well as by using single high $\pt$ particles and simulated jets embedded into real \PbPb\ events and reconstructed with the anti-\kt\ jet finder. The influence of a low transverse momentum cut-off on particles used in the jet reconstruction is quantified by varying the minimum track $\pt$ between 0.15\,GeV/$c$ and 2\,GeV/$c$. For embedded jets reconstructed from charged particles with $\pt > 0.15$\,GeV/$c$, the
uncertainty in the reconstructed jet transverse momentum due to the heavy-ion background is measured to be 11.3\,GeV/$c$ (standard deviation) for the 10\% most central \PbPb\ collisions, slightly larger than the value of 11.0\,GeV/$c$ measured using the unbiased method. For a higher particle transverse momentum threshold of 2\,GeV/$c$, which will generate a stronger bias towards hard fragmentation in the jet finding process, the standard deviation of the fluctuations in the reconstructed jet transverse momentum is reduced to 4.8-5.0\,GeV/$c$ for the 10\% most central events. 
A non-Gaussian tail of the momentum uncertainty is observed and its impact on the reconstructed jet spectrum is evaluated for varying particle momentum thresholds, by folding the measured fluctuations with steeply falling spectra.

\end{abstract}
\end{frontmatter}

}
\setcounter{page}{2}






%

\section{Introduction}

High energy heavy-ion collisions explore strongly interacting matter
under extreme conditions of energy density, where lattice QCD
predicts a phase transition to a new state of matter above a critical
value of about 1\,GeV/fm$^3$ \cite{Karsch:2003jg}. In this new state,
called the Quark-Gluon Plasma (QGP), quarks and gluons rather than
hadrons are expected to be the dominant degrees of freedom over length
scales larger than that of a nucleon.
Experiments studying the collision of heavy nuclei at high energy at
both the Relativistic Heavy Ion Collider (RHIC)
\cite{Adams:2005dq,Adcox:2004mh,Arsene:2004fa,Back:2004je}, and
recently at the Large Hadron Collider (LHC)
\cite{Aamodt:2010pb,Aamodt:2010pa,Chatrchyan:2011pb,Aad:2010bu}, have made
several key observations that point to the formation of a hot, dense
and strongly coupled system, possibly the QGP.

Hard (large momentum transfer $Q^2$) probes are well calibrated 
tools to study the properties of the matter created in such collisions.  The
scattered partons generated in a hard momentum exchange 
are created in the initial stages of the heavy-ion
collision, with production rates that are calculable using perturbative QCD, which can be compared
to the same measurements in proton-proton collisions. The scattered partons 
then propagate through the medium, where their fragmentation into observed jets of 
hadrons is expected to be modified relative to the vacuum case by interactions with the medium 
(\emph{jet quenching}) \cite{Gyu90a,Baier:1996sk}. 
This modification of parton fragmentation provides sensitive
observables to study properties of the created matter.

Jet quenching has been observed at RHIC 
\cite{Adcox:2001jp,Adler:2003qi,Ada03b,Adams:2003im} and at the LHC
\cite{Aamodt:2010jd} via the measurement of high \pt\ hadron inclusive
production and correlations, which are observed to be strongly
suppressed in central \AA\ collisions compared to a scaled \pp\
reference. These high \pt\ hadron observables have been the major tool for
measuring the energy loss of hard scattered partons and thereby the
properties of the medium, but they provide only indirect and biased
information on the parton evolution in the medium.
The aim of full jet reconstruction is to measure jet modifications due to energy loss 
in an unbiased way \cite{Salgado:2003rv,Alessandro:2006yt}.  Already first measurements of 
reconstructed jets in heavy-ion collisions at the LHC showed an energy imbalance between 
back-to-back dijets, which is attributed to jet quenching \cite{Aad:2010bu,Chatrchyan:2011sx}. 

Jet reconstruction in the complex environment of a
heavy-ion collision requires a quantitative understanding of
background-induced fluctuations of the measured jet signal and the
effects of the underlying heavy-ion event on the jet finding process
itself.  Here, region-to-region background fluctuations are the main
source of jet energy or momentum uncertainty and can have a large impact on jet
structure observables, such as the fraction of energy inside the jet
core or the shape of the jet, and will distort the measured jet energy
balance even in the absence of medium effects \cite{Cacciari:2011tm}.

In this paper the measurement of jet transverse momentum fluctuations
due to the background in heavy-ion collisions is reported  and its
sources are identified, based on jet reconstruction using charged
particles with varying minimum track $\pt$.
For this purpose three methods are employed to probe the
measured \PbPb\ events:
fixed area (\emph{rigid}) cones placed randomly in the acceptance, the
simulation of high-$\pt$ single
tracks or full jets from \pp\ collisions. 
Rigid cones enable the identification of 
contributions to the fluctuations in an unbiased fashion, while 
single tracks and embedded jets explore the interplay between the jet
finding process, the underlying
event, and the jet fragmentation pattern.

\section{Detector Description and Track Selection}

The data presented here were collected by the ALICE experiment
\cite{Aamodt:2008zz} in the first \PbPb\ run of the LHC in November 2010, at
a collision energy of $\sqrt{s_{_{\rm NN}}} = 2.76$\,TeV.
This analysis is based on minimum-bias events, triggered
by two forward VZERO counters and the Silicon Pixel Detector (SPD)\cite{Collaboration:2010ys}. 
A description of the minimum-bias trigger can be found in \cite{Aamodt:2010pb}.
The VZERO trigger counters are forward scintillator detectors covering a pseudo-rapidity range of $2.8 < \eta < 5.1$
(V0A) and $-3.7 < \eta < -1.7$ (V0C). 
The sum of VZERO amplitudes is also used as a measure of event centrality
\cite{Aamodt:2010cz}.
The SPD consists of two silicon pixel layers at a radial distance to the beam line of $r = 3.9$~cm and 
$r = 7.6$~cm.

To ensure a uniform track acceptance in pseudo-rapidity $\eta$, only events
whose primary vertex lies within $\pm8$\,cm from the center of the detector along the beam
line are used, resulting in 13.3\,M minimum-bias \PbPb\ events for this analysis.

Charged particle tracking is carried out using the Time Projection Chamber (TPC)
\cite{Alme:2010ke} and the Inner Tracking System (ITS) \cite{Collaboration:2010ys}, 
located in the central barrel of the ALICE experiment within a $0.5$\,T solenoidal
magnetic field and covering the full azimuth within pseudo-rapidity $|\eta| < 0.9$.
The ITS consists of six cylindrical layers of silicon detectors, with distances from the beam-axis between
$r = 3.9$\,cm and $r = 43$\,cm. The ITS layers measure track points close to the primary vertex, with 
the two innermost layers (SPD) providing a precise measurement of the primary vertex position.
The TPC, a cylindrical drift detector surrounding the ITS, is the
main tracking detector in ALICE. The TPC inner radius is $85$\,cm and the outer radius is $247$\,cm, with
longitudinal coverage $-250 < z < 250$\,cm. It
provides a uniformly high tracking efficiency for charged particles. 
The high precision of the ITS and the large radial lever arm of the TPC provide a good momentum resolution for 
combined (\emph{global}) tracks.
%

%


For this dataset the ITS has significantly non-uniform efficiency as a
function of azimuthal angle $\phi$ and pseudo-rapidity $\eta$. In
order to obtain high and uniform tracking efficiency together with
good momentum resolution, two different track populations are
utilized: (i) tracks containing at least one space-point reconstructed
in one of the two innermost layers of the ITS (78\% of all accepted
tracks), and (ii) accepted tracks that lack the position information
close to the beam-line. Here, the primary vertex is added to the fit
of the track which modifies the reconstructed curvature of the charged
track in the magnetic field. Since the majority of the tracks
originates from the primary vertex this constraint improves the
momentum resolution. Both track types have transverse momentum
resolution of $\sigma(\pt)/\pt \approx 1\%$ at $1$\,GeV/$c$. For the
majority of tracks the resolution at $\pt = 50$\,GeV/$c$ is
$\sigma(\pt)/\pt \approx 10\%$, only tracks having fewer than three
reconstructed space points in the ITS (about 6\% of the total
population) have a resolution at $50$\,GeV/$c$ of $\sigma(\pt)/\pt
\approx 20\%$.

Tracks are accepted for $\pt > 0.15$\,GeV/$c$ and $|\eta| < 0.9$. The
tracking efficiency at $\pt = 0.15$\,GeV/$c$ is 50\%, increasing to 90\%
at 1\,GeV/$c$ and above. Tracks with measured $\pt >100$\,GeV/$c$ are
accepted at the tracking stage, but jets containing them are rejected
from the analysis to reduce the influence of fake tracks and limited tracking resolution at very high $\pt$.

\section{Jet Reconstruction and Background Subtraction}

\begin{figure}
\begin{center}
{\includegraphics[width=0.48\textwidth]{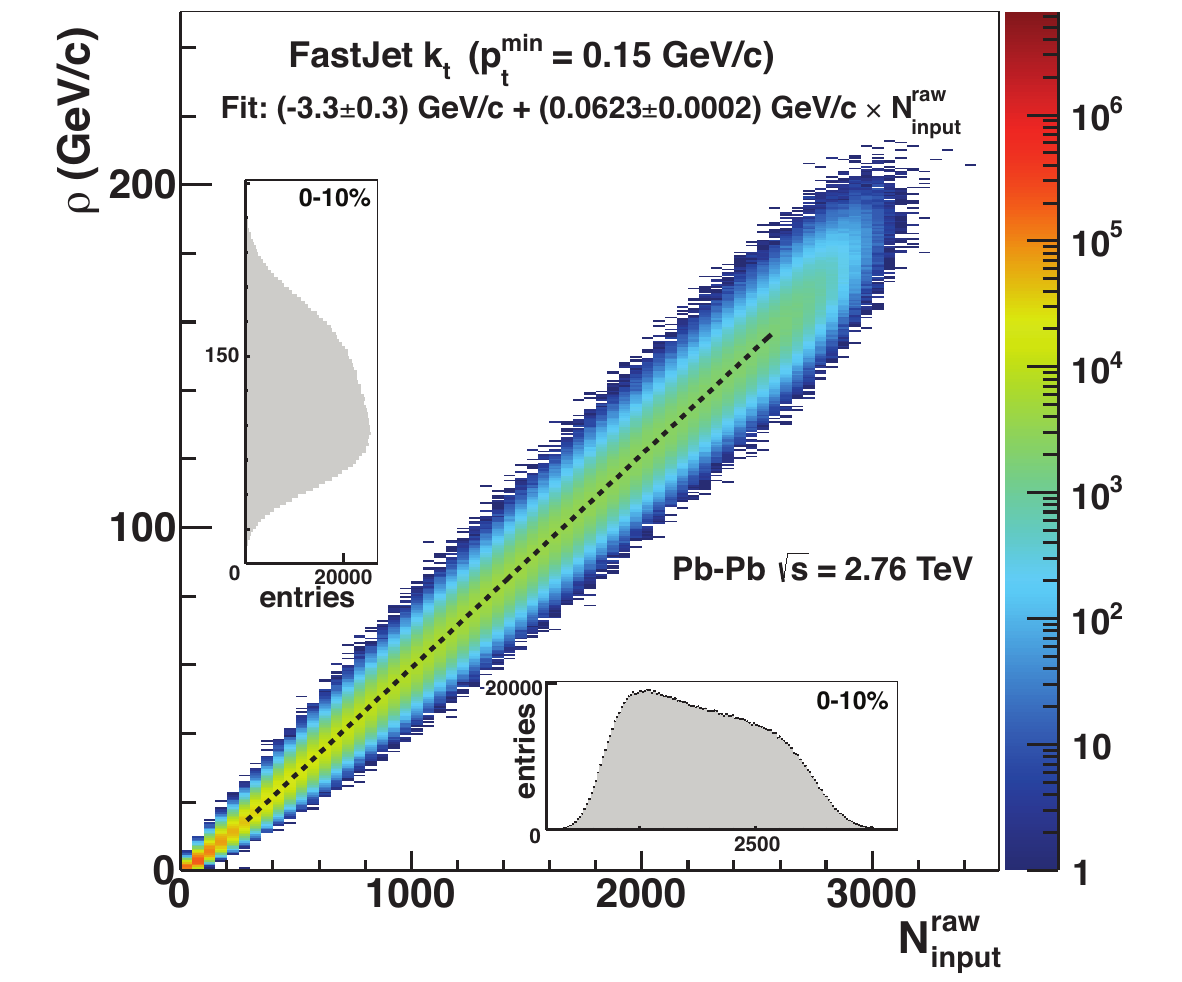}}
\end{center}
\caption{Dependence of charged particle background \pt\ density $\rho$ on uncorrected 
    multiplicity of tracks used for jet finding ($|\eta| < 0.9$). The dotted line is a linear fit to the
    centroids in each multiplicity bin. The insets show the projected distributions of
    $\rho$ and raw multiplicity for the 10\% most central events. 
    \label{fig:MultVsRho}
}
\end{figure}

Charged particle jet reconstruction and estimation of the background employ
the sequential recombination algorithms {anti-\kt} and  {\kt} from
 the FastJet package \cite{Cacciari:2005hq}.
The clustering starts with the list of tracks that
satisfy the quality, acceptance, and \pt-cuts, with no pre-clustering or
grouping of tracks. 
A list of jet
candidates (anti-\kt) or clusters (\kt) is generated, with direction
and  transverse momentum given
by the \pt-weighted average of ($\eta,\phi$) of the individual constituents
and the scalar sum of their \pt, respectively. 
The distance parameter that determines the terminating condition for the
clustering is chosen as $R = $ 0.4, which is a common value for reconstruction
of jets in heavy-ion collisions
\cite{Alessandro:2006yt,Aad:2010bu,Salur:2008hs}.
As proposed in \cite{Cacciari:2007fd}, the clusters found by the \kt\
algorithm are used to estimate the event-wise background $\pt$-density per
unit area, $\rho$, defined as the median value of the ratio $\pt^{\rm
  rec}/A^{\rm rec}$  for all
considered \kt-clusters. $A^{\rm rec}$ is
the area of the reconstructed cluster in the ($\eta,\phi$)-plane calculated by the active ghost area method of
FastJet \cite{Cacciari:2008gn}, with a ghost area of 0.005.
To minimize the influence of the track acceptance interval on  $\rho$, only
reconstructed clusters  with $|\eta| < 0.5$ have been used. In
addition, the two clusters with the largest $\pt^{\rm rec}$ (leading) in the full acceptance of $|\eta| <
0.9$ are excluded from the calculation of the median to further reduce
the influence of true jets on the background estimate \cite{Cacciari:2010te}.
The jet population reconstructed by the {anti-\kt} algorithm is used
as signal jets. Their $\pt$ is corrected for the
background \pt-density in each event using the jet area $A^{\rm jet,rec}$ with $\pt^{\rm
  jet} = \pt^{\rm jet,rec} - \rho \cdot A^{\rm jet,rec}$. Signal jets
are only considered for  $|\eta| < 0.5$.

The average transverse momentum of tracks
$\langle\pt\rangle$ and the total charged multiplicity are global observables
that are closely related to the value of $\rho$, though the
determination of $\rho$ uses varying phase-space intervals 
(with typical areas in the ($\eta,\phi$)-plane of $\pi R^2$)
and suppresses hard jet contributions by using the median of the distribution.
Figure~\ref{fig:MultVsRho} shows the correlation between $\rho$ and the
uncorrected multiplicity of tracks with $|\eta| < 0.9$. 
The linear increase corresponds to an
uncorrected $\langle\pt\rangle$ of about 0.7\,GeV/$c$ per accepted charged track.
%
Both $\langle \pt\rangle$ and multiplicity, and thus also the
background \pt\ density, strongly depend on the minimum \pt\ threshold (\ptmin)
applied for tracks used as input to the jet finding.
To minimize the bias on jet fragmentation, a value of
$\ptmin\ = 0.15$\,GeV/$c$ is preferred. In addition, $\ptmin\ =$ 1 and 2\,GeV/$c$ are
investigated  to facilitate comparisons to other experiments and to Monte-Carlo generators in a
region of constant and high tracking efficiency. 
The mean $\rho$ over all events and its standard deviation is given for
different $\ptmin$ and two centralities in Table~\ref{tab:rho}. As
expected, the mean
background \pt\ density decreases for larger \ptmin, for central
collisions and $\ptmin\ = $2\,GeV/$c$ it is reduced by an order of magnitude.
As one can see in the insets
of Figure~\ref{fig:MultVsRho} and the standard deviation in Table~\ref{tab:rho}, the spread of $\rho$ for the 10\%
most central events is considerable, underlining the importance of the
event-by-event background subtraction.

\begin{table}
\begin{center}
 \begin{tabular}{c|c|c}
\ptmin\ & $\langle \rho \rangle$ & $\sigma(\rho)$\\
(GeV/$c$) & (GeV/$c$) & (GeV/$c$) \\
\hline
  \multicolumn{3}{l}{0-10\%}\\  
\hline
 0.15  &$138.32 \pm 0.02$ &  $18.51 \pm 0.01$  \\
1.00  &  $59.30 \pm 0.01$ &  $9.27 \pm 0.01$  \\
2.00  & $12.28 \pm 0.01$ &  $3.29 \pm 0.01$  \\
\hline
  \multicolumn{3}{l}{50-60\%}\\  
\hline
0.15  &  $12.05 \pm 0.01$ &  $3.41 \pm 0.01$ \\ 
1.00  &   $4.82 \pm 0.01$ &  $1.77 \pm 0.01$ \\
2.00  &  $4.41 \pm 0.05$ &  $0.92 \pm 0.04$  \\
\end{tabular}
\end{center}
\caption{
\label{tab:rho}
Average and standard deviation  of the event-wise charged
particle \pt\ density $\rho$ for three choices of minimum particle
$\pt$ and two centrality bins. The quoted uncertainties are purely statistical.}
\end{table}

\section{Sources of Background Fluctuations}

To study the sources of background fluctuations in an unbiased way
that is not influenced by a particular choice of jet finder, a single
rigid cone with radius $R = 0.4$ is placed in each reconstructed event
at random $\phi$ and $\eta$, with centroid lying within $|\eta| <
0.5$. The background fluctuations are characterized by calculating the
difference of the scalar sum of all track $\pt$ in the cone and the expected background:
\begin{equation}
\label{eq:deltaRC}
\delta{\pt} = \sum_{i}p_{\rm t,i} - A \cdot \rho,
\end{equation}
where $A = \pi R^2$.

The rigid random cone (RC) area and position are not influenced by the event, so that it provides
a sampling of the event structure at the typical scale of a jet, but independent of biases
induced by the choice of a particular jet
finding algorithm. The RC measurements will be compared to the embedding of
jet-like objects, that is
directly relevant to the measurement of the inclusive jet spectrum
with a specific choice of jet finder.

%

To characterize the $\delta\pt$ distribution the standard
deviation, $\sigma(\delta\pt)$, is
utilized. In addition, a Gaussian distribution with mean $\mu^{\rm LHS}$
and standard deviation $\sigma^{\rm LHS}$ is iteratively fit to the
distribution within $[\mu^{\rm LHS} - 3
\sigma^{\rm LHS}, \mu^{\rm LHS}  + 0.5 \sigma^{\rm LHS}]$, i.e. to the left-hand-side. 
The $\sigma^{\rm LHS}$ of the fit provides the lower limit on the
magnitude of the fluctuations and is used to characterize shape
differences between the positive and negative tails of the
distribution, by extrapolating the Gaussian distribution to positive
$\delta\pt$.

Various sources contribute to background
fluctuations in a heavy-ion event, including:
(i) random,
uncorrelated (Poissonian) fluctuations of particle number and
momentum; (ii) region-to-region
correlated variations of the momentum density, induced
by detector effects, e.g. a non-uniform efficiency,
and by the heavy-ion collision itself, e.g. by variation of the
eccentricity of the nuclear overlap for collisions with finite impact parameter.

%

%
%

\begin{figure}
\begin{center}
{\includegraphics[width=0.48\textwidth]{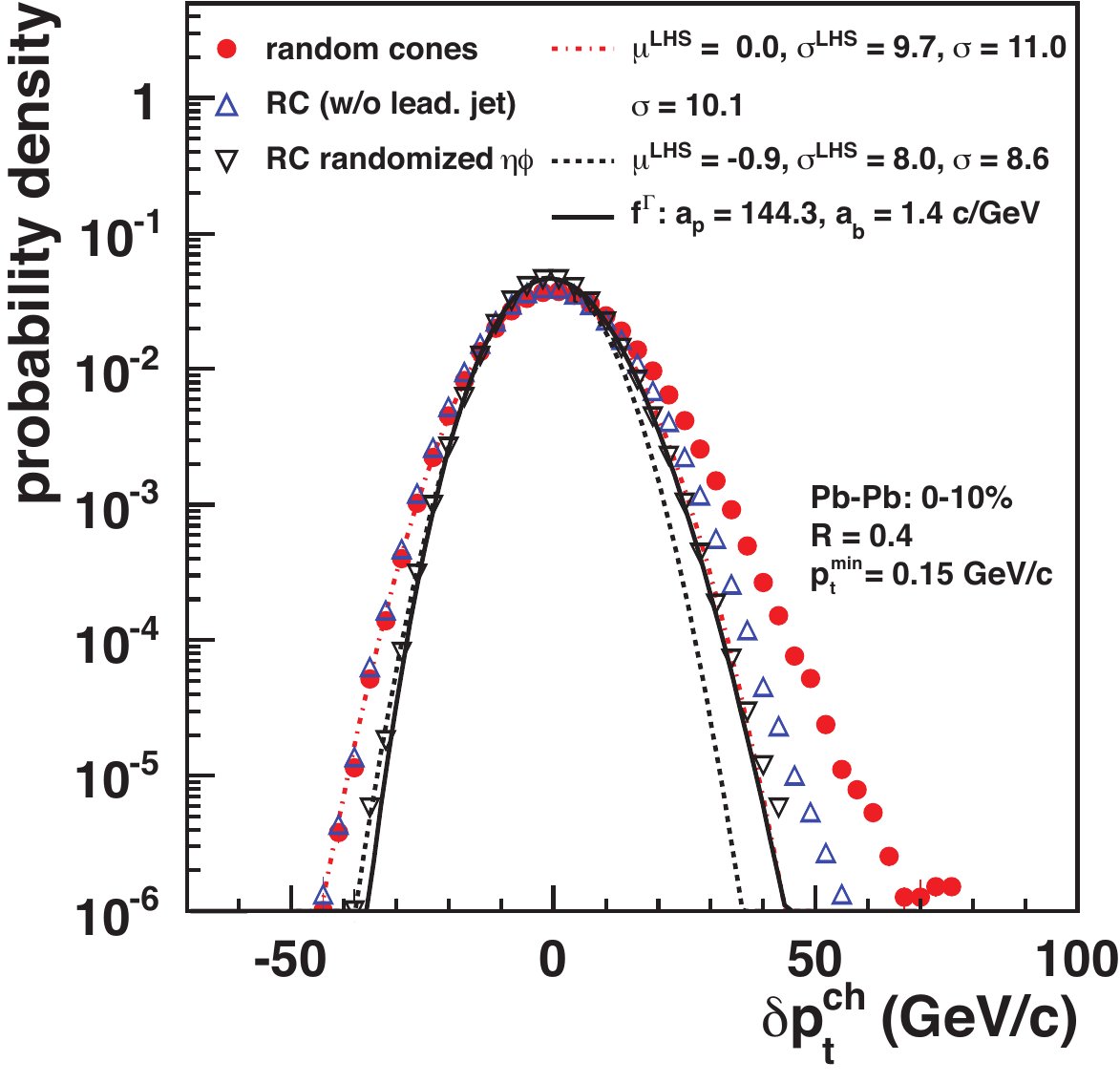}}
\end{center}
 \caption{  $\delta{\pt}$ of random cones in  the 10\% most central
    \PbPb\ events for the three types of random cone
    probes with $\ptmin = 0.15$\,GeV/$c$. A Gaussian fit to the
    left-hand-side and its extrapolation to positive $\delta\pt$ are
    shown for measured \PbPb\ events and for randomized \PbPb\ events ($\mu^{\rm LHS}$ and
    $\sigma^{\rm LHS}$ in GeV/$c$). The solid line is a fit to the $\delta\pt$
    distribution for the randomized events with a $\Gamma$
    distribution shifted to zero (Equation~\ref{eq:gamma}) as
    approximation for the shape in case of  independent particle emission.
   \label{fig:deltaptRC0-10}
}
\end{figure}

The measured $\delta\pt$ distribution for random cones in the 10\%
most central \PbPb\ events is shown in Figure~\ref{fig:deltaptRC0-10}.
The distribution is peaked near zero, illustrating the agreement of
the background estimate via \kt-clusters and that due to random
sampling of the event. 
The distribution exhibits an asymmetric shape with a tail to the
right-hand-side of the distribution, which is also reflected in the
difference between the standard deviation of the full distribution of
$\sigma(\delta\pt)$ = 11.0\,GeV/$c$ and the Gaussian width of the
left-hand-side $\sigma^{\rm LHS}(\delta\pt) =$ 9.6\,GeV/$c$.
%

To further differentiate random and correlated sources of
fluctuations, three variations of the random cone method are employed:
(i) sampling of measured \PbPb\ events, (ii) sampling of measured
\PbPb\ events, but avoiding overlap with the leading jet candidate in
the event after background subtraction by requiring a distance $D =
1.0$ in ($\eta,\phi$) between the random cone direction and the jet
axis, and (iii) sampling of \PbPb\ events in which the ($\eta$,$\phi$)
direction of the tracks has been randomized within the acceptance,
which destroys all correlations in the event.
Figure~\ref{fig:deltaptRC0-10} shows that when avoiding the leading
jet candidate to suppress upward fluctuations, e.g. due to a hard process, the
tail to the right-hand-side is already significantly reduced.
 
Note that, even for the case of purely statistical fluctuations, the
distribution is not expected to be symmetric or to follow a Gaussian
shape on the right-hand-side, since the shapes of the underlying
single particle $\pt$ and multiplicity distributions are not
Gaussian. In the case of uncorrelated particle emission a
$\Gamma$-distribution provides a more accurate description of the
event-wise $\left< \pt \right>$ fluctuations \cite{Tannenbaum:2001gs}.
This also holds for $\delta{\pt}$ distributions, which are similar to
a measurement of $\left<\pt\right>$ fluctuations in a limited interval
of phase space. Taking into account the subtraction of the average
background the functional form of the probability distribution of
$\delta\pt$ for independent particle emission can be written:
\begin{equation}
\label{eq:gamma}
 f^{\Gamma}(\delta\pt) = A \cdot a_{b} /\Gamma(a_{p}) \cdot (a_b \delta\pt+
 a_p)^{a_p-1} \cdot e^{-(a_b \delta\pt + a_p)}.
\end{equation}
This corresponds to a $\Gamma$-distribution with mean shifted from $a_p/a_b$ 
to zero and standard deviation $\sigma = \sqrt{a_p}/a_b$.
As seen in Figure~\ref{fig:deltaptRC0-10} this functional form provides a good approximation of the $\delta\pt$
distribution for randomized events, corresponding to uncorrelated
emission.
In this case the distribution is also narrower on the left-hand-side.  
This points to the presence of correlated region-to-region fluctuations in
addition to purely statistical fluctuations and those expected from
hard processes.

 %
  %
  %

\begin{figure}
\begin{center}
{\includegraphics[width=0.48\textwidth]{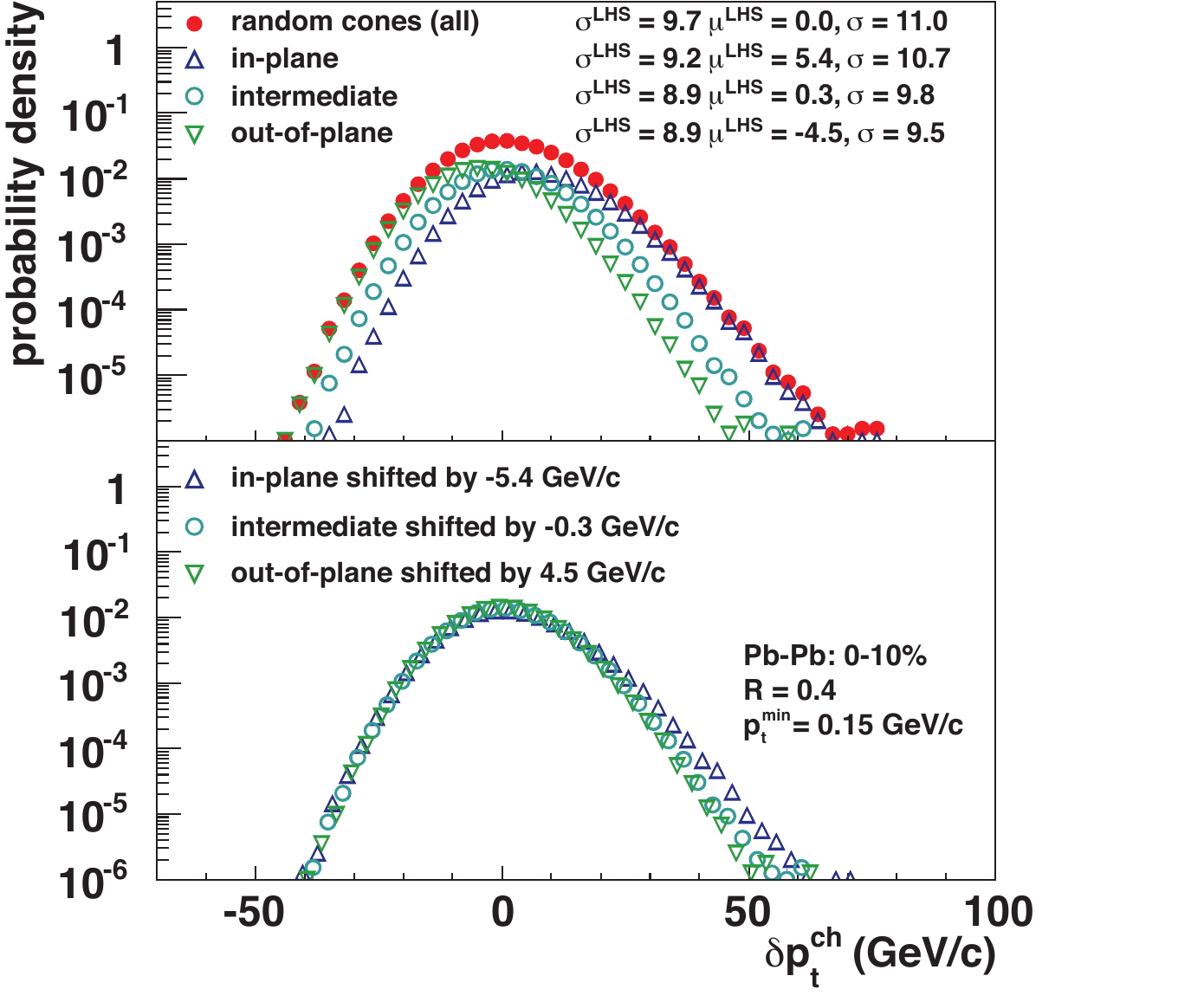}}
\end{center}
 \caption{$\delta\pt$ distribution for random cones, averaged over the full azimuth 
   and separated for three bins of random cone azimuthal orientations with respect to the measured
   event plane. In the bottom panel the
   distributions have been shifted to zero using the mean of the
   left-hand-side Gaussian fit ($\mu^{\rm LHS}$).
 \label{fig:deltaptEP0-10}}
\end{figure}

One source of region-to-region variation in the background \pt\
density is the initial anisotropy of the nuclear overlap for finite
impact parameter collisions, which
translates via the collective expansion of the medium into an anisotropy
in momentum space \cite{Aamodt:2010pa,Ollitrault:1992bk} with respect
to the symmetry plane of the collision. The event plane direction can be calculated using 
the azimuthal distribution of all accepted tracks within each event, which is dominated by soft particle 
production. The final state hadron azimuthal distribution with respect to the reaction plane
of the event,
is characterized by a Fourier expansion where the leading term is
the second moment, called elliptic flow $v_2$. In addition to the
geometry driven even harmonics (mainly $v_2$), odd flow components
(e.g. $v_3$) driven by intial
state fluctuations can modify the azimuthal distribution within the
event \cite{:2011vk}.

To explore the effect on background fluctuations of azimuthal
orientation relative to the reaction plane, the $\delta\pt$
distribution from random cone sampling is studied as a function of
the azimuthal orientation of the cone axis, $\phi$, relative to the reconstructed event
plane orientation, $\psi_{\rm RP}$.
Three bins are chosen; the out-of-plane orientation where the azimuthal
angle between the reconstructed event plane and the cone axis is $> 60^\circ$,
the in-plane orientation where this angle is $< 30^\circ$, and the
intermediate orientation where the angle is between $30$ and $60^\circ$.
The distributions of $\delta\pt$ for random cones averaged over the
full azimuth and for the three different orientations are shown in
Figure~\ref{fig:deltaptEP0-10}. It can be seen that, for out-of-plane
cones, the most probable background \pt\ density is smaller by
almost 5\,GeV/$c$ relative to the azimuthally averaged estimate of
$\rho$, with opposite effect in-plane.
This shift scales with the average flow and the background \pt\ density for
a given centrality ($\propto\,v_{2}\cdot\rho$), and is seen to be 
sizable in central events, though discrimination of the event plane
orientations  is limited by finite event-plane resolution \cite{Bielcikova:2003ku} and possible
biases due to hard jets. The decreasing 
width of left-hand-side Gaussian is qualitatively consistent with the
expectation from reduced particle number fluctuations out-of-plane
compared to in-plane. For a visual
comparison of their shape, the distributions have been
shifted such that the centroid of the left-hand-side Gaussian fit is
zero (see Figure~\ref{fig:deltaptEP0-10}).
Notably, the left-hand-side of the distribution appears similar for all
orientations of the random cones to the event-plane. The random cones distributed in-plane show
a more pronounced tail to the right-hand-side, compared to
out-of-plane. This may point to a dependence of the jet spectrum on
the orientation relative to the reaction plane, though further
systematic studies are needed to assess biases in the event plane
determination due to jet production and possible auto-correlations.
For the measurement of the inclusive jet spectrum the correction via
an event-plane dependent $\rho$ will reduce the influence of even flow
components on the average reconstructed jet momentum, but its systematic precision
is limited by the finite event-plane resolution.


\begin{figure}
\begin{center}
\includegraphics[width=0.48\textwidth]{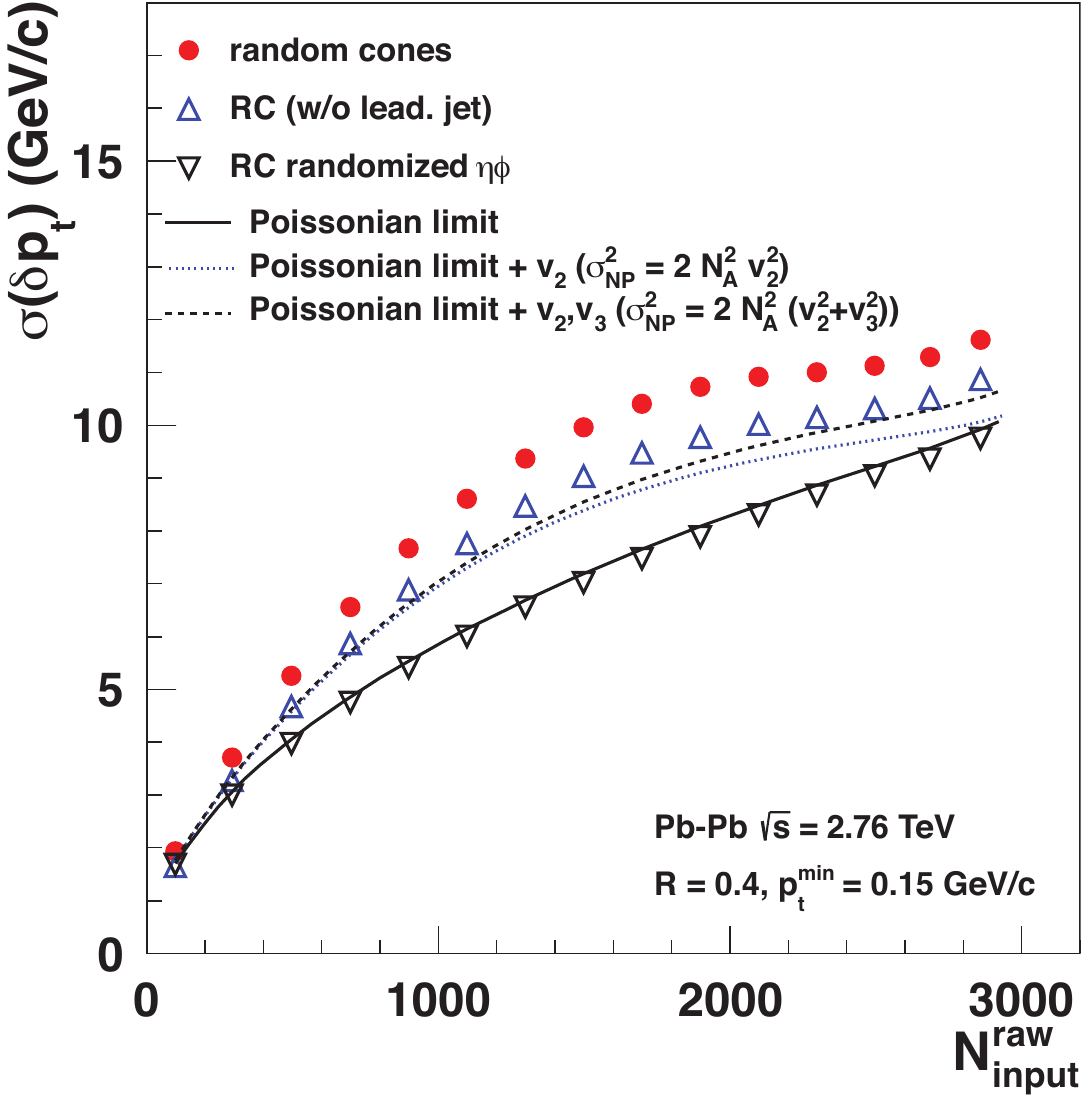}
\end{center}
 \caption{
   \label{fig:sigmaVsMult}
   Dependence of the standard deviation of the $\delta\pt$
   distributions on uncorrected charged particle multiplicity, compared to the limit
   derived from the measured track \pt\ spectrum (Equation~\ref{eq:stat})
   and from additional elliptic and triangular flow contributions (Equation~\ref{eq:statcorr}).
  $R = 0.4$, $\ptmin = 0.15$\,GeV/$c$. 
}
\end{figure}

The width of the $\delta\pt$ distribution due to purely random fluctuations can be estimated from 
the measured single particle \pt\ spectrum via \cite{Alessandro:2006yt}:
\begin{equation}
\sigma(\delta{\pt}) =  \sqrt{\NA \cdot \sigma^2(\pt) + \NA \cdot \langle  \pt \rangle^2}.
\label{eq:stat}
\end{equation}
Here, $\NA$ is the expected number of tracks in the cone area $A$ for a
given event centrality or multiplicity class, $\langle \pt \rangle$ the average $\pt$ and
$\sigma(\pt)$ the standard deviation of the track \pt\ spectrum. 
Local variations of the average multiplicity, $\langle \pt \rangle$,
or $\sigma(\pt)$, lead to additional fluctuations.  These
region-to-region variations can be induced e.g. by (mini-)jets, where
the particle \pt\ spectrum is considerably harder than for the global
event average, and by collective flow.  Uncorrelated non-Poissonian (NP)
fluctuations can be added to Equation~\ref{eq:stat} knowing their
standard deviation, e.g. for
additional region-to-region variation of the average multiplicity:
\begin{equation}
\sigma(\delta{\pt}) = \sqrt{\NA \cdot \sigma^2(\pt) + \left(\NA + \sigma^2_{\rm
    NP}{(\NA)}\right)  \cdot \langle \pt \rangle ^2}.
\label{eq:statcorr}
\end{equation}

Figure~\ref{fig:sigmaVsMult} shows the comparison of
the multiplicity dependence of $\sigma(\delta\pt)$ for the three different types of random cones.
The distribution of purely statistical fluctuations given by
Equation~\ref{eq:stat} well describes the randomized events. 
Also shown are two parameterizations following
Equation~\ref{eq:statcorr}.
Additional multiplicity fluctuations due to elliptic flow are
approximated from the \pt-integrated $v_2$ values measured by ALICE
for different centralities \cite{Aamodt:2010pa} as $\sigma_{\rm
  NP}^2(\NA) \approx 2 v_2^2 \NA^2$. This approximate inclusion of
$v_2$-effects accounts qualitatively for the larger fluctuations
in mid central collisions compared to the randomized events and the deviation from a
$\sqrt{N}$-increase. The random cone sampling with an anti-bias on the
leading jet has, by construction, a reduced standard
deviation and is close to the parameterization of elliptic
flow. Taking into account also region-to-region fluctuations from triangular
flow, $v_3$, is of particular importance in central events where it
reaches a similar magnitude as $v_2$ \cite{:2011vk}. The contribution
of $v_3$ can be added in quadrature  ($\sigma_{\rm
  NP}^2(\NA) \approx 2 \NA^2 (v^2_2 + v_3^2)$) since the second
  and the third harmonic are not correlated via a common plane of
  symmetry \cite{:2011vk}, for simplicity $v_3$ has been approximated by a
  constant value of $v_3 = 2.4\%$. As expected, the inclusion of $v_3$
  can account partially for the difference to the randomized event in
  the most central events. In the comparison one has to consider
  that in practice the contribution from hard processes to the
  right-hand-side tail cannot be cleanly separated from (soft) upward
  multiplicity fluctuations induced by flow. In addition, the
  approximate description of flow effects following  Equation~\ref{eq:statcorr}
  does not take into account any flow-correlated changes of $\langle \pt \rangle$ and
  $\sigma(\pt)$. 


The track reconstruction efficiency affects the total
multiplicity and the shape of the measured \pt-spectrum at low
\pt. Using Equation~\ref{eq:stat}, the change of the uncorrelated
fluctuations due to finite efficiency can be estimated from the efficiency corrected
\pt-spectrum in each centrality bin. This procedure suggests that, for \ptmin =
0.15\,GeV/$c$, there is an increase of the standard deviation by 5.4--6.0\%,
depending on centrality.  The complete correction requires the
knowledge of all correlations within the heavy-ion event and is beyond the scope of the present study.



%
%

\section{Background Fluctuations in Jet Reconstruction}

The measured jet spectrum in heavy-ion collisions is affected over the
entire \pt\ range by background fluctuations, especially due to the
large and asymmetric tail 
towards positive $\delta\pt$. 
For the measurement of
the inclusive jet cross section, background fluctuations can only be
corrected on a statistical basis via unfolding.  Such background
fluctuations are evaluated using embedding and reconstruction of a
probe with identical jet algorithm and parameters as those applied to
the data analysis, to account for the jet-finder-specific response to
the heavy-ion background.

In the present study, two probes are embedded into the \PbPb\
events measured by ALICE: (i) single high-\pt\ tracks at various
$\pt$, and (ii) \pp\ jet events generated using PYTHIA \cite{Sjostrand:2006za}
followed by a detailed simulation of the full detector response. 
%
Jet candidates are reconstructed from the event using the anti-\kt\
algorithm with $R = 0.4$ and matched to the
embedded probe, by either finding the single track in it, or by
requiring that the \pt\ of the embedded tracks within the
reconstructed jet sum up to at least 50\% of the original probe jet
transverse momentum ($\pt^{\rm probe}$).
The difference between the reconstructed, background subtracted jet and the embedded probe is
  then given, similar to
Equation~\ref{eq:deltaRC}, by \cite{Cacciari:2010te,Jacobs:2010wq}:
\begin{equation}
 \delta{\pt} = \pt^{\rm jet,rec} - A^{\rm jet,rec} \cdot \rho - \pt^{\rm probe}.
\end{equation}
The response may depend on the jet finder, its settings, and the
properties of the embedded probe, such as $\pt^{\rm probe}$, area, and
fragmentation pattern. In particular the insensitivity to the
latter is essential for a robust and unbiased reconstruction of jets
in heavy-ion collisions, where the fragmentation pattern is
potentially modified relative to that in \pp\ collisions, and is
indeed the observable of interest.

\begin{figure}
\begin{center}
{\includegraphics[width=0.48\textwidth]{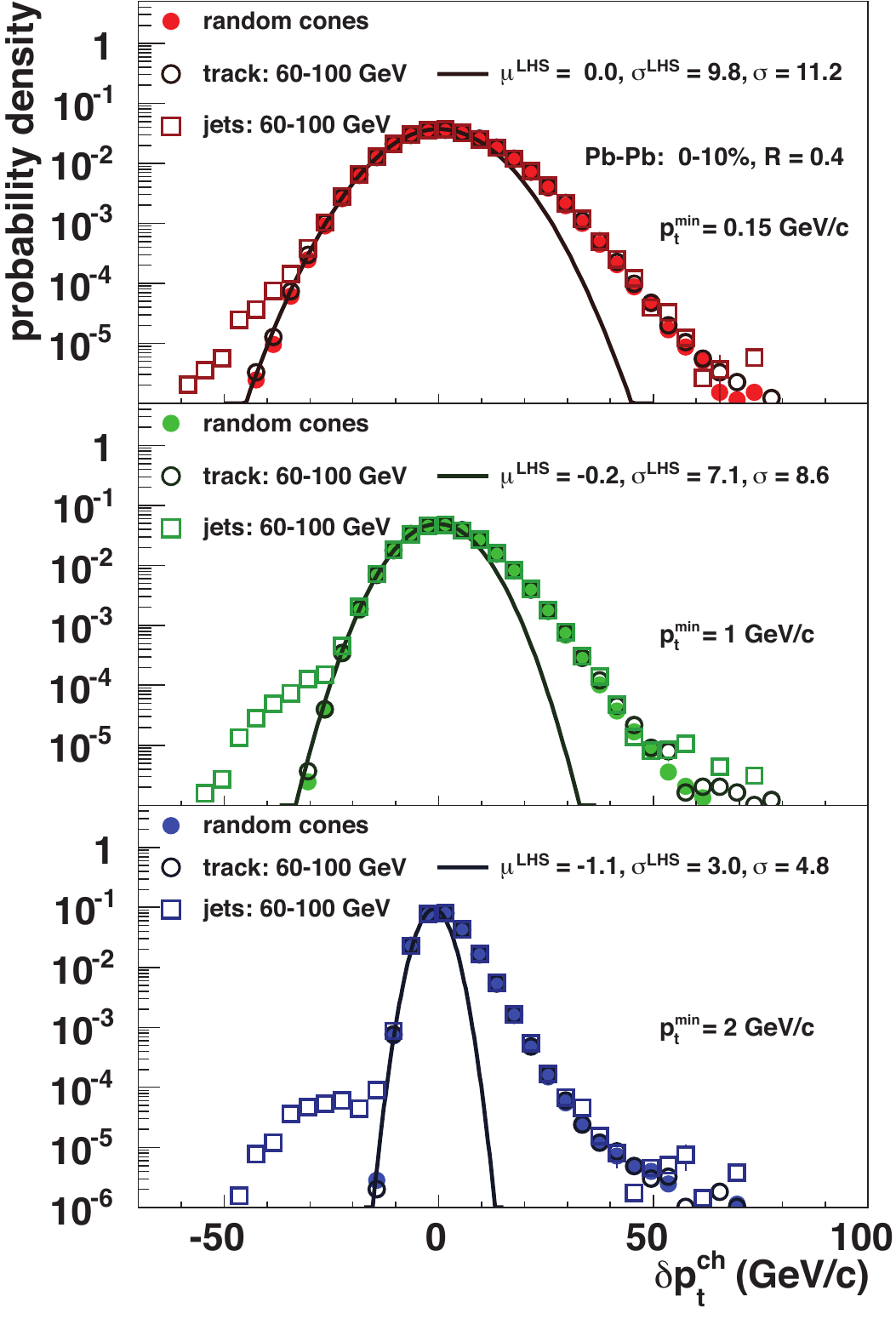}}
\end{center}
  \caption{  $\delta{\pt}$ distribution of charged particles for jet
    reconstruction with the three methods in the 10\% most central
    \PbPb\ events for $\ptmin = 0.15$\,GeV/$c$, 1\,GeV/$c$, and 2\,GeV/$c$.
   \label{fig:deltaptAll0-10}
}
\end{figure}

The $\delta\pt$ distributions measured for each of the methods are
shown in Figure~\ref{fig:deltaptAll0-10}. Here, the focus is on high
$\pt^{\rm probe}$ ($> 60$\,GeV/$c$), where the efficiency of matching
the embedded probe to the reconstructed jet approaches unity. 
The results are very similar to the random cone method, including the
presence of an asymmetric tail to the right-hand-side of the
distribution. The standard deviations, however, show a small increase
compared to the random cone method, which is largest for jet embedding
(see Table~\ref{tab:sigmaCentral}). The increase may be due to
sensitivity of the jet finder to back-reaction, e.g. the stability of
the probe area and jet direction after embedding.  The single particle
embedding can be considered as extreme fragmentation leading to rather
rigid cones with stable area $\pi R^2$, while in the case of true
\pp-jets the probe and reconstructed area may differ, depending on the
fragmentation pattern.  In addition the finite jet area resolution due
to the size of the ghost area has to be taken into account
\cite{Cacciari:2010te}. With a ghost area of 0.005 a compromise
between reasonable jet area resolution and computing time and memory
consumption was chosen.  In the case of track embedding at high $\pt$,
the jet area resolution fully accounts for the difference of 200\,MeV/$c$ observed
in the standard deviation.


The broadening of the $\delta\pt$ distribution for jets with \ptmin =
0.15\,GeV/$c$, as seen in Figure~\ref{fig:deltaptAll0-10}, has been
investigated more closely. The additional left-hand-side
structure is caused by probe jets with large area ($A^{\rm probe} >
0.6$) that are split in the heavy-ion event into two separate objects
of smaller size. Jets with a large area ($A > 0.6$) are only formed by
the anti-$\kt$ algorithm in exceptional cases, where there are two
hard cores at distance close to $R$ \cite{Cacciari:2008gp}.  It is
also seen in Figure~\ref{fig:deltaptAll0-10} that, with increasing
$\ptmin$, the deviations on the left-hand-side in the case of jet
embedding become more pronounced, suggesting that the jet-splitting is
an effect of hard fragmentation. 

In the determination of $\delta\pt$ fluctuations as described above,
the probes have been embedded into an event population recorded with a
minimum-bias trigger. However, the requirement of a hard process
biases the population towards more central (small impact parameter)
collisions, due to nuclear geometry. Correction of this effect for
centrality bins of 10\% width, generates a negligible increase of the
fluctuations ($< 0.1$\,GeV/$c$).
The full centrality dependence of the fluctuations is given via the
standard deviation of the distributions and for different \ptmin\ cuts
in Table~\ref{tab:sigmacent}.

The increase of the $\ptmin$ cut on the input tracks for jet finding
reduces the background fluctuations, due to the smaller
influence of statistical and soft region-to-region fluctuations. This
is observed in Figure~\ref{fig:deltaptAll0-10} when the $\ptmin$ is
varied from $0.15$ to $2$\,GeV/$c$.  A $\ptmin$ of $2$\,GeV/$c$
reduces the standard deviation by more than a factor of two compared to
$0.15$\,GeV/$c$. Soft region-to-region fluctuations that dominate the
left-hand-side of the distribution are reduced by a factor of three
(see Table~\ref{tab:sigmaCentral}). A high $\ptmin$ significantly
reduces the impact of fluctuations in the jet spectrum (see
Table~\ref{tab:yieldChange}). However, it may also introduce a bias in
the jet reconstruction towards hard fragmentation.



To estimate the influence of the observed fluctuations on the jet
measurement, a power law spectrum starting at $\pt = 4$\,GeV/$c$ has
been folded with a Gaussian of width $\sigma^{\rm Gauss} =
11$\,GeV/$c$ ($5$\,GeV/$c$) and with the measured $\delta\pt$
distributions for \ptmin\ = 0.15\,GeV/$c$ (2\,GeV/$c$).  The 
yield increase relative to the unsmeared spectrum in one high
$\pt$-bin for the most central collisions is given in
Table~\ref{tab:yieldChange}. For the different probes, they agree
within the uncertainties given by the statistical fluctuations in the
tails of the $\delta\pt$-distributions; about a factor of ten increase for
\ptmin\ = 0.15\,GeV/$c$ and a 30\% effect for \ptmin\ = 2\,GeV/$c$. 
Minor differences in the
standard deviation as well as the left-hand-side differences have no
sizable effect on the spectral shape after folding. The difference
between smearing with the full $\delta\pt$ and with a Gaussian
distribution illustrates the strong influence of the right-hand-side
tail, which must be taken into account in the analysis of background
fluctuation effects on jet reconstruction. The extracted values
naturally depend on the choice of the input spectrum, so, in addition
to the power law, a jet spectrum for \pp\ collisions at $\sqrt{s_{\rm
    NN}}$ = 2.76\,TeV extracted from PYTHIA simulations has been
used. These studies indicate that the increase of yield due to
background fluctuations falls below 50\% for reconstructed charged
jets in the region of $\pt \approx 100 \pm 15$\,GeV/$c$ ($60 \pm
10$\,GeV/$c$) for $\ptmin = 0.15$\,GeV/$c$ (2\,GeV/$c$). Repeating the
exercise with a Gaussian smearing of a $\sigma^{\rm Gauss} =
11$\,GeV/$c$ and with the $\delta\pt$ distribution of random cones
avoiding the leading jet for $\ptmin =
0.15$\,GeV/$c$ leads, as expected, to a reduced influence of the
tail. Here, the relative yield increase falls below 50\% in the range
of $\pt \approx 75 \pm 10$\,GeV/$c$. The employed
input spectra do not consider the geometrical limitation on the
number of jets that can be reconstructed within the acceptance for a single event
\cite{deBarros:2011ph}. This effect also limits the extraction of jet spectra at lower
\pt\ via unfolding.


\begin{table}
\begin{center}
\begin{tabular}{>{\small}l|>{\small}c|>{\small}c|>{\small}c}
&$\sigma$ (GeV/$c$) &  $\sigma^{\rm LHS}$ (GeV/$c$) & $\mu^{\rm LHS}$ (GeV/$c$)  \\
\hline
  \multicolumn{4}{l}{\small $\ptmin = 0.15$ GeV/$c$}\\
\hline
 random cones & $10.98 \pm 0.01$ & $9.65 \pm 0.02$ &  $-0.04 \pm 0.03$ \\
  track emb. & $11.19 \pm 0.01$ & $9.80 \pm 0.02$ &  $0.00 \pm 0.03$ \\
 jet emb. &  $11.34 \pm 0.02$ & $9.93 \pm 0.06$ &  $0.06 \pm 0.09$ \\
\hline
  \multicolumn{4}{l}{\small $\ptmin = 1$ GeV/$c$}\\
\hline
random cones &  $8.50 \pm 0.01$ & $7.08 \pm 0.01$ &  $-0.22 \pm 0.02$ \\
 track emb. & $8.61 \pm 0.01$ & $7.11 \pm 0.01$ &  $-0.25 \pm 0.02$ \\
 jet emb. &$8.78 \pm 0.02$ & $7.25 \pm 0.04$ &  $-0.08 \pm 0.08$ \\
\hline
  \multicolumn{4}{l}{\small $\ptmin = 2$ GeV/$c$}\\
\hline
random cones&  $4.82 \pm 0.01$ & $3.41 \pm 0.01$ &  $-0.01 \pm 0.01$ \\
 track emb. &  $4.88 \pm 0.01$ & $3.05 \pm 0.01$ &  $-0.92 \pm 0.01$ \\
 jet emb. & $5.03 \pm 0.01$ & $3.52 \pm 0.01$ &  $0.01 \pm 0.02$ 
\end{tabular}
\end{center}
\caption{Background fluctuations in central events.
Comparison of the Gaussian fit to the left-hand-side of the
  $\delta\pt$-distributions and the standard deviation in central \PbPb\
  collisions for the three different methods and for the three
  \ptmin-cuts. The quoted uncertainties are purely statistical.
  \label{tab:sigmaCentral}
}
\end{table}

\begin{table}
\begin{center}
\begin{tabular}{>{\small}c|>{\small}c|>{\small}c|>{\small}c}
&  \multicolumn{3}{c}{ \small$\ptmin$ (GeV/$c$)}\\
              & 0.15  & 1.0 & 2.0 \\
\hline
Centrality Class &  \multicolumn{3}{c}{\small $\sigma(\delta\pt)$ (GeV/$c$)}\\
\hline
0-10\%   & $11.19 \pm 0.01$   &$8.61 \pm 0.01$ &$4.88 \pm 0.01$ \\
10-20\% & $10.19 \pm 0.01$   &$7.67 \pm 0.01$ & $4.29 \pm 0.01$\\
20-30\% & $8.46 \pm 0.01$     &$6.35 \pm 0.01$ &$3.58 \pm 0.01$ \\
30-40\% & $6.51 \pm 0.01$     &$4.93 \pm 0.01$ &$2.68 \pm 0.01$ \\
40-50\% & $4.71 \pm 0.01$     &$3.63 \pm 0.01$ &$1.95 \pm 0.01$ \\
50-60\% & $3.28 \pm 0.01$     &$2.61 \pm 0.01$ &$1.41 \pm 0.01$ \\
60-70\% & $2.22 \pm 0.01$     &$1.70 \pm 0.01$ &$0.95 \pm 0.01$ \\
70-80\% & $1.48 \pm 0.01$     &$1.01 \pm 0.01$ &$0.62 \pm 0.01$ \\
\end{tabular}
\end{center}
\caption{Centrality dependence of fluctuations.
Standard deviation of $\delta\pt$ distributions and statistical uncertainty for different centrality
  bins and \ptmin\ cuts using the track embedding probe.
\label{tab:sigmacent}}
\end{table}

\begin{table}
\begin{center}
\begin{tabular}{>{\small}c|>{\small}c|>{\small}c|>{\small}c}
$f(\pt)$ folded with &  \multicolumn{3}{c}{\small relative yield for
  $\pt =
  60-68$\,GeV/$c$}\\
\hline
 $\delta\pt$  &  RC  &  tracks &  jets \\
\cline{2-4}
$\ptmin = 0.15$\,GeV/$c$ & $9.8 \pm 1.7$& $11.4 \pm 1.1$  & $10.9 \pm 3.4$ \\
$\ptmin = 2$\,GeV/$c$ & $1.30 \pm 0.02$ & $1.31 \pm 0.02$ & $1.65 \pm 0.25$\\ 
\hline
Gauss &  \multicolumn{3}{c}{}\\
$\sigma = 11$\,GeV/$c$ & \multicolumn{3}{c}{\small $1.82 \pm 0.04$}\\ 
$\sigma = 5$\,GeV/$c$ & \multicolumn{3}{c}{\small $1.05 \pm 0.01$} \\
\end{tabular}
\end{center}
\caption{Yield modification for power law spectrum.
Relative yield in the bin $\pt =
  60-68$\,GeV/$c$ for a power law spectrum ($f(\pt) = 0.7/(0.7
  + \pt^5)$ and $\pt > 4$\,GeV/$c$), folded with the
  different $\delta\pt$ distributions for 0-10\% centrality and with
  a Gaussian, where the width is similar to the standard
  deviation of the $\delta\pt$ distributions.
\label{tab:yieldChange}}
\end{table}


\section{Summary}

The first detailed study of event background fluctuations for jet
reconstruction using charged particles in \PbPb\ collisions at the LHC has been presented. The standard deviation of the
fluctuations in the 10\% most central events is $\sigma = (10.98 \pm 0.01)$
\,GeV/$c$ within a rigid cone of $R = 0.4$ and for a low \pt\ cut-off
of 0.15\,GeV/$c$. It
has been shown that the non-statistical sources of fluctuations are
driven in part by the anisotropy of the particles emitted from the
collision (elliptic and triangular flow). The variation of multiplicity in different
orientations with respect to the event plane, induces shifts in
the background-subtracted jet $\pt$ even for central
\PbPb-collisions.


The anti-{\kt} jet finder response for charged particle jet
reconstruction has a modest dependence on the method used to
characterize the fluctuations. For embedded, simulated \pp-jets the standard deviation increases
to $(11.34 \pm 0.02)$\,GeV/$c$.  In addition, certain rare
fragmentation patterns in \pp\ are likely to be split in the heavy-ion
environment leading to minor effects in the background response.  The
observed differences between the two types of embedded probes
(namely single tracks and \pp\ jets) do not indicate a strong sensitivity of the reconstructed anti-\kt\
jet spectrum on fragmentation. The case of a strong broadening of the
jet due to medium effects has not been considered here.

The use of reconstructed charged particles down to $\ptmin =
0.15$\,GeV/$c$ allows a comparison of the impact of
background fluctuations with a minimal bias on hard fragmentation in
jet finding to the case with increased bias ($\ptmin \ge
1$\,GeV/$c$). The observed reduction of the standard deviation to  $\sigma =
(4.82 \pm 0.01)$\,GeV/$c$ for the unbiased sampling and $\ptmin = 2$\,GeV/$c$  is
driven by the smaller number fluctuations and the reduced influence of
soft region-to-region fluctuations. 

The asymmetric shape of the $\delta \pt$ distribution with a tail
towards positive fluctuations has a large impact on the jet
measurement, compared to purely Gaussian fluctuations, though the role
of signal jets contributing to the tail has to be considered. 
Using different
assumptions on the shape of the true jet spectrum it is found that for
$\ptmin = 0.15$\,GeV/$c$ fluctuations can have a large influence on the charged jet
yield for transverse momenta up to $100 \pm 15$\,GeV/$c$. 

%




\ifthenelse{\equal{\publisher}{cern}}{
\section*{Acknowledgements}
The ALICE collaboration would like to thank all its engineers and technicians for their invaluable contributions to the construction of the experiment and the CERN accelerator teams for the outstanding performance of the LHC complex.
\\
The ALICE collaboration acknowledges the following funding agencies for their support in building and
running the ALICE detector:
 \\
Calouste Gulbenkian Foundation from Lisbon and Swiss Fonds Kidagan, Armenia;
 \\
Conselho Nacional de Desenvolvimento Cient\'{\i}fico e Tecnol\'{o}gico (CNPq), Financiadora de Estudos e Projetos (FINEP),
Funda\c{c}\~{a}o de Amparo \`{a} Pesquisa do Estado de S\~{a}o Paulo (FAPESP);
 \\
National Natural Science Foundation of China (NSFC), the Chinese Ministry of Education (CMOE)
and the Ministry of Science and Technology of China (MSTC);
 \\
Ministry of Education and Youth of the Czech Republic;
 \\
Danish Natural Science Research Council, the Carlsberg Foundation and the Danish National Research Foundation;
 \\
The European Research Council under the European Community's Seventh Framework Programme;
 \\
Helsinki Institute of Physics and the Academy of Finland;
 \\
French CNRS-IN2P3, the `Region Pays de Loire', `Region Alsace', `Region Auvergne' and CEA, France;
 \\
German BMBF and the Helmholtz Association;
\\
General Secretariat for Research and Technology, Ministry of
Development, Greece;
\\
Hungarian OTKA and National Office for Research and Technology (NKTH);
 \\
Department of Atomic Energy and Department of Science and Technology of the Government of India;
 \\
Istituto Nazionale di Fisica Nucleare (INFN) of Italy;
 \\
MEXT Grant-in-Aid for Specially Promoted Research, Ja\-pan;
 \\
Joint Institute for Nuclear Research, Dubna;
 \\
National Research Foundation of Korea (NRF);
 \\
CONACYT, DGAPA, M\'{e}xico, ALFA-EC and the HELEN Program (High-Energy physics Latin-American--European Network);
 \\
Stichting voor Fundamenteel Onderzoek der Materie (FOM) and the Nederlandse Organisatie voor Wetenschappelijk Onderzoek (NWO), Netherlands;
 \\
Research Council of Norway (NFR);
 \\
Polish Ministry of Science and Higher Education;
 \\
National Authority for Scientific Research - NASR (Autoritatea Na\c{t}ional\u{a} pentru Cercetare \c{S}tiin\c{t}ific\u{a} - ANCS);
 \\
Federal Agency of Science of the Ministry of Education and Science of Russian Federation, International Science and
Technology Center, Russian Academy of Sciences, Russian Federal Agency of Atomic Energy, Russian Federal Agency for Science and Innovations and CERN-INTAS;
 \\
Ministry of Education of Slovakia;
 \\
Department of Science and Technology, South Africa;
 \\
CIEMAT, EELA, Ministerio de Educaci\'{o}n y Ciencia of Spain, Xunta de Galicia (Conseller\'{\i}a de Educaci\'{o}n),
CEA\-DEN, Cubaenerg\'{\i}a, Cuba, and IAEA (International Atomic Energy Agency);
 \\
Swedish Reseach Council (VR) and Knut $\&$ Alice Wallenberg Foundation (KAW);
 \\
Ukraine Ministry of Education and Science;
 \\
United Kingdom Science and Technology Facilities Council (STFC);
 \\
The United States Department of Energy, the United States National
Science Foundation, the State of Texas, and the State of Ohio.

}

\ifthenelse{\equal{\publisher}{elsevier}}{
}

\ifthenelse{\equal{\publisher}{jhep}}{

}

\ifthenelse{\equal{\publisher}{elsevier}}{
\bibliographystyle{elsarticle-num}
}

\ifthenelse{\equal{\publisher}{cern}}{
\bibliographystyle{unsrt}
}

\ifthenelse{\equal{\publisher}{jhep}}{
\bibliographystyle{JHEP}
}

\bibliography{fluctuations_master}

\ifthenelse{\equal{\publisher}{cern}}{
\newpage
\appendix
\section{The ALICE Collaboration}
\label{app:collab}

\begingroup
\small
\begin{flushleft}
B.~Abelev\Irefn{org1234}\And
J.~Adam\Irefn{org1274}\And
D.~Adamov\'{a}\Irefn{org1283}\And
A.M.~Adare\Irefn{org1260}\And
M.M.~Aggarwal\Irefn{org1157}\And
G.~Aglieri~Rinella\Irefn{org1192}\And
A.G.~Agocs\Irefn{org1143}\And
A.~Agostinelli\Irefn{org1132}\And
S.~Aguilar~Salazar\Irefn{org1247}\And
Z.~Ahammed\Irefn{org1225}
% ...
\renewcommand\labelenumi{\textsuperscript{\theenumi}~}
\section*{Affiliation notes}
\renewcommand\theenumi{\roman{enumi}}
\begin{Authlist}
\item \Adef{0}Deceased
\item \Adef{M.V.Lomonosov Moscow State University, D.V.Skobeltsyn Institute of Nuclear Physics, Moscow, Russia}Also at: M.V.Lomonosov Moscow State University, D.V.Skobeltsyn Institute of Nuclear Physics, Moscow, Russia
\item \Adef{Institute of Nuclear Sciences, Belgrade, Serbia}Also at: "Vin\v{c}a" Institute of Nuclear Sciences, Belgrade, Serbia
\end{Authlist}
\section*{Collaboration Institutes}
\renewcommand\theenumi{\arabic{enumi}~}
\begin{Authlist}
\item \Idef{org1279}Benem\'{e}rita Universidad Aut\'{o}noma de Puebla, Puebla, Mexico
\item \Idef{org1220}Bogolyubov Institute for Theoretical Physics, Kiev, Ukraine
\item \Idef{org1262}Budker Institute for Nuclear Physics, Novosibirsk, Russia
\item \Idef{org1292}California Polytechnic State University, San Luis Obispo, California, United States
\item \Idef{org14939}Centre de Calcul de l'IN2P3, Villeurbanne, France
\item \Idef{org1197}Centro de Aplicaciones Tecnol\'{o}gicas y Desarrollo Nuclear (CEADEN), Havana, Cuba
\item \Idef{org1242}Centro de Investigaciones Energ\'{e}ticas Medioambientales y Tecnol\'{o}gicas (CIEMAT), Madrid, Spain
\item \Idef{org1244}Centro de Investigaci\'{o}n y de Estudios Avanzados (CINVESTAV), Mexico City and M\'{e}rida, Mexico
\item \Idef{org1335}Centro Fermi -- Centro Studi e Ricerche e Museo Storico della Fisica ``Enrico Fermi'', Rome, Italy
\item \Idef{org17347}Chicago State University, Chicago, United States
\item \Idef{org1118}China Institute of Atomic Energy, Beijing, China
\item \Idef{org1288}Commissariat \`{a} l'Energie Atomique, IRFU, Saclay, France
\item \Idef{org1294}Departamento de F\'{\i}sica de Part\'{\i}culas and IGFAE, Universidad de Santiago de Compostela, Santiago de Compostela, Spain
\item \Idef{org1106}Department of Physics Aligarh Muslim University, Aligarh, India
\item \Idef{org1121}Department of Physics and Technology, University of Bergen, Bergen, Norway
\item \Idef{org1162}Department of Physics, Ohio State University, Columbus, Ohio, United States
\item \Idef{org1300}Department of Physics, Sejong University, Seoul, South Korea
\item \Idef{org1268}Department of Physics, University of Oslo, Oslo, Norway
\item \Idef{org1132}Dipartimento di Fisica dell'Universit\`{a} and Sezione INFN, Bologna, Italy
\item \Idef{org1315}Dipartimento di Fisica dell'Universit\`{a} and Sezione INFN, Trieste, Italy
\item \Idef{org1145}Dipartimento di Fisica dell'Universit\`{a} and Sezione INFN, Cagliari, Italy
\item \Idef{org1270}Dipartimento di Fisica dell'Universit\`{a} and Sezione INFN, Padova, Italy
\item \Idef{org1285}Dipartimento di Fisica dell'Universit\`{a} `La Sapienza' and Sezione INFN, Rome, Italy
\item \Idef{org1154}Dipartimento di Fisica e Astronomia dell'Universit\`{a} and Sezione INFN, Catania, Italy
\item \Idef{org1290}Dipartimento di Fisica `E.R.~Caianiello' dell'Universit\`{a} and Gruppo Collegato INFN, Salerno, Italy
\item \Idef{org1312}Dipartimento di Fisica Sperimentale dell'Universit\`{a} and Sezione INFN, Turin, Italy
\item \Idef{org1103}Dipartimento di Scienze e Tecnologie Avanzate dell'Universit\`{a} del Piemonte Orientale and Gruppo Collegato INFN, Alessandria, Italy
\item \Idef{org1114}Dipartimento Interateneo di Fisica `M.~Merlin' and Sezione INFN, Bari, Italy
\item \Idef{org1237}Division of Experimental High Energy Physics, University of Lund, Lund, Sweden
\item \Idef{org1192}European Organization for Nuclear Research (CERN), Geneva, Switzerland
\item \Idef{org1227}Fachhochschule K\"{o}ln, K\"{o}ln, Germany
\item \Idef{org1122}Faculty of Engineering, Bergen University College, Bergen, Norway
\item \Idef{org1136}Faculty of Mathematics, Physics and Informatics, Comenius University, Bratislava, Slovakia
\item \Idef{org1274}Faculty of Nuclear Sciences and Physical Engineering, Czech Technical University in Prague, Prague, Czech Republic
\item \Idef{org1229}Faculty of Science, P.J.~\v{S}af\'{a}rik University, Ko\v{s}ice, Slovakia
\item \Idef{org1184}Frankfurt Institute for Advanced Studies, Johann Wolfgang Goethe-Universit\"{a}t Frankfurt, Frankfurt, Germany
\item \Idef{org1215}Gangneung-Wonju National University, Gangneung, South Korea
\item \Idef{org1212}Helsinki Institute of Physics (HIP) and University of Jyv\"{a}skyl\"{a}, Jyv\"{a}skyl\"{a}, Finland
\item \Idef{org1203}Hiroshima University, Hiroshima, Japan
\item \Idef{org1329}Hua-Zhong Normal University, Wuhan, China
\item \Idef{org1254}Indian Institute of Technology, Mumbai, India
\item \Idef{org1266}Institut de Physique Nucl\'{e}aire d'Orsay (IPNO), Universit\'{e} Paris-Sud, CNRS-IN2P3, Orsay, France
\item \Idef{org1277}Institute for High Energy Physics, Protvino, Russia
\item \Idef{org1249}Institute for Nuclear Research, Academy of Sciences, Moscow, Russia
\item \Idef{org1320}Nikhef, National Institute for Subatomic Physics and Institute for Subatomic Physics of Utrecht University, Utrecht, Netherlands
\item \Idef{org1250}Institute for Theoretical and Experimental Physics, Moscow, Russia
\item \Idef{org1230}Institute of Experimental Physics, Slovak Academy of Sciences, Ko\v{s}ice, Slovakia
\item \Idef{org1127}Institute of Physics, Bhubaneswar, India
\item \Idef{org1275}Institute of Physics, Academy of Sciences of the Czech Republic, Prague, Czech Republic
\item \Idef{org1139}Institute of Space Sciences (ISS), Bucharest, Romania
\item \Idef{org27399}Institut f\"{u}r Informatik, Johann Wolfgang Goethe-Universit\"{a}t Frankfurt, Frankfurt, Germany
\item \Idef{org1185}Institut f\"{u}r Kernphysik, Johann Wolfgang Goethe-Universit\"{a}t Frankfurt, Frankfurt, Germany
\item \Idef{org1177}Institut f\"{u}r Kernphysik, Technische Universit\"{a}t Darmstadt, Darmstadt, Germany
\item \Idef{org1256}Institut f\"{u}r Kernphysik, Westf\"{a}lische Wilhelms-Universit\"{a}t M\"{u}nster, M\"{u}nster, Germany
\item \Idef{org1246}Instituto de Ciencias Nucleares, Universidad Nacional Aut\'{o}noma de M\'{e}xico, Mexico City, Mexico
\item \Idef{org1247}Instituto de F\'{\i}sica, Universidad Nacional Aut\'{o}noma de M\'{e}xico, Mexico City, Mexico
\item \Idef{org23333}Institut of Theoretical Physics, University of Wroclaw
\item \Idef{org1308}Institut Pluridisciplinaire Hubert Curien (IPHC), Universit\'{e} de Strasbourg, CNRS-IN2P3, Strasbourg, France
\item \Idef{org1182}Joint Institute for Nuclear Research (JINR), Dubna, Russia
\item \Idef{org1143}KFKI Research Institute for Particle and Nuclear Physics, Hungarian Academy of Sciences, Budapest, Hungary
\item \Idef{org18995}Kharkiv Institute of Physics and Technology (KIPT), National Academy of Sciences of Ukraine (NASU), Kharkov, Ukraine
\item \Idef{org1199}Kirchhoff-Institut f\"{u}r Physik, Ruprecht-Karls-Universit\"{a}t Heidelberg, Heidelberg, Germany
\item \Idef{org20954}Korea Institute of Science and Technology Information
\item \Idef{org1160}Laboratoire de Physique Corpusculaire (LPC), Clermont Universit\'{e}, Universit\'{e} Blaise Pascal, CNRS--IN2P3, Clermont-Ferrand, France
\item \Idef{org1194}Laboratoire de Physique Subatomique et de Cosmologie (LPSC), Universit\'{e} Joseph Fourier, CNRS-IN2P3, Institut Polytechnique de Grenoble, Grenoble, France
\item \Idef{org1187}Laboratori Nazionali di Frascati, INFN, Frascati, Italy
\item \Idef{org1232}Laboratori Nazionali di Legnaro, INFN, Legnaro, Italy
\item \Idef{org1125}Lawrence Berkeley National Laboratory, Berkeley, California, United States
\item \Idef{org1234}Lawrence Livermore National Laboratory, Livermore, California, United States
\item \Idef{org1251}Moscow Engineering Physics Institute, Moscow, Russia
\item \Idef{org1140}National Institute for Physics and Nuclear Engineering, Bucharest, Romania
\item \Idef{org1165}Niels Bohr Institute, University of Copenhagen, Copenhagen, Denmark
\item \Idef{org1109}Nikhef, National Institute for Subatomic Physics, Amsterdam, Netherlands
\item \Idef{org1283}Nuclear Physics Institute, Academy of Sciences of the Czech Republic, \v{R}e\v{z} u Prahy, Czech Republic
\item \Idef{org1264}Oak Ridge National Laboratory, Oak Ridge, Tennessee, United States
\item \Idef{org1189}Petersburg Nuclear Physics Institute, Gatchina, Russia
\item \Idef{org1170}Physics Department, Creighton University, Omaha, Nebraska, United States
\item \Idef{org1157}Physics Department, Panjab University, Chandigarh, India
\item \Idef{org1112}Physics Department, University of Athens, Athens, Greece
\item \Idef{org1152}Physics Department, University of Cape Town, iThemba LABS, Cape Town, South Africa
\item \Idef{org1209}Physics Department, University of Jammu, Jammu, India
\item \Idef{org1207}Physics Department, University of Rajasthan, Jaipur, India
\item \Idef{org1200}Physikalisches Institut, Ruprecht-Karls-Universit\"{a}t Heidelberg, Heidelberg, Germany
\item \Idef{org1325}Purdue University, West Lafayette, Indiana, United States
\item \Idef{org1281}Pusan National University, Pusan, South Korea
\item \Idef{org1176}Research Division and ExtreMe Matter Institute EMMI, GSI Helmholtzzentrum f\"ur Schwerionenforschung, Darmstadt, Germany
\item \Idef{org1334}Rudjer Bo\v{s}kovi\'{c} Institute, Zagreb, Croatia
\item \Idef{org1298}Russian Federal Nuclear Center (VNIIEF), Sarov, Russia
\item \Idef{org1252}Russian Research Centre Kurchatov Institute, Moscow, Russia
\item \Idef{org1224}Saha Institute of Nuclear Physics, Kolkata, India
\item \Idef{org1130}School of Physics and Astronomy, University of Birmingham, Birmingham, United Kingdom
\item \Idef{org1338}Secci\'{o}n F\'{\i}sica, Departamento de Ciencias, Pontificia Universidad Cat\'{o}lica del Per\'{u}, Lima, Peru
\item \Idef{org1146}Sezione INFN, Cagliari, Italy
\item \Idef{org1115}Sezione INFN, Bari, Italy
\item \Idef{org1313}Sezione INFN, Turin, Italy
\item \Idef{org1133}Sezione INFN, Bologna, Italy
\item \Idef{org1155}Sezione INFN, Catania, Italy
\item \Idef{org1316}Sezione INFN, Trieste, Italy
\item \Idef{org1286}Sezione INFN, Rome, Italy
\item \Idef{org1271}Sezione INFN, Padova, Italy
\item \Idef{org1322}Soltan Institute for Nuclear Studies, Warsaw, Poland
\item \Idef{org1258}SUBATECH, Ecole des Mines de Nantes, Universit\'{e} de Nantes, CNRS-IN2P3, Nantes, France
\item \Idef{org1304}Technical University of Split FESB, Split, Croatia
\item \Idef{org1168}The Henryk Niewodniczanski Institute of Nuclear Physics, Polish Academy of Sciences, Cracow, Poland
\item \Idef{org17361}The University of Texas at Austin, Physics Department, Austin, TX, United States
\item \Idef{org1173}Universidad Aut\'{o}noma de Sinaloa, Culiac\'{a}n, Mexico
\item \Idef{org1296}Universidade de S\~{a}o Paulo (USP), S\~{a}o Paulo, Brazil
\item \Idef{org1149}Universidade Estadual de Campinas (UNICAMP), Campinas, Brazil
\item \Idef{org1239}Universit\'{e} de Lyon, Universit\'{e} Lyon 1, CNRS/IN2P3, IPN-Lyon, Villeurbanne, France
\item \Idef{org1205}University of Houston, Houston, Texas, United States
\item \Idef{org20371}University of Technology and Austrian Academy of Sciences, Vienna, Austria
\item \Idef{org1222}University of Tennessee, Knoxville, Tennessee, United States
\item \Idef{org1310}University of Tokyo, Tokyo, Japan
\item \Idef{org1318}University of Tsukuba, Tsukuba, Japan
\item \Idef{org21360}Eberhard Karls Universit\"{a}t T\"{u}bingen, T\"{u}bingen, Germany
\item \Idef{org1225}Variable Energy Cyclotron Centre, Kolkata, India
\item \Idef{org1306}V.~Fock Institute for Physics, St. Petersburg State University, St. Petersburg, Russia
\item \Idef{org1323}Warsaw University of Technology, Warsaw, Poland
\item \Idef{org1179}Wayne State University, Detroit, Michigan, United States
\item \Idef{org1260}Yale University, New Haven, Connecticut, United States
\item \Idef{org1332}Yerevan Physics Institute, Yerevan, Armenia
\item \Idef{org15649}Yildiz Technical University, Istanbul, Turkey
\item \Idef{org1301}Yonsei University, Seoul, South Korea
\item \Idef{org1327}Zentrum f\"{u}r Technologietransfer und Telekommunikation (ZTT), Fachhochschule Worms, Worms, Germany
\end{Authlist}
\endgroup

}

\end{document}